\documentclass[12pt]{article} 
\usepackage{graphicx}
\newcommand{\be}{\begin{equation}}
\newcommand{\ee}{\end{equation}}
\newcommand{\bear}{\begin{eqnarray}}
\newcommand{\eear}{\end{eqnarray}}
\newcommand{\ba}{\begin{array}}
\newcommand{\ea}{\end{array}}

\begin{document}

\begin{titlepage}
\vfill
\begin{flushright}
{\normalsize IC/2009/XXX}\\
{\normalsize arXiv:XXXX.XXXX[hep-th]}\\
\end{flushright}

\vfill
\begin{center}
{\Large\bf Holographic chiral shear waves from anomaly  }

\vskip 0.3in

{ Bindusar Sahoo\footnote{\tt bsahoo@ictp.it}, Ho-Ung
Yee\footnote{\tt hyee@ictp.it}}

\vskip 0.15in

 {\it ICTP, High Energy, Cosmology and Astroparticle Physics,} \\
{\it Strada Costiera 11, 34014, Trieste, Italy}
\\[0.3in]

{\normalsize  2009}

\end{center}

\vfill

\begin{abstract}

We study dispersion relations of hydrodynamic waves of hot N=4 SYM plasma at strong coupling with a finite U(1) R-charge chemical potential via holography.
We first provide complete equations of motion of linearized fluctuations out of a charged AdS black-hole
background according to their helicity, and observe that helicity $\pm 1$ transverse shear modes receive a new parity-odd contribution from the 5D Chern-Simons term, which is dual to  4D $U(1)^3$ anomaly. We present a systematic solution of the
helicity $\pm 1$ wave equations in long wave-length expansion, and obtain the corresponding dispersion relations. The results depend on the sign of helicity, which may be called chiral shear waves.

\end{abstract}

\vfill

\end{titlepage}
\setcounter{footnote}{0}

\baselineskip 18pt \pagebreak
\renewcommand{\thepage}{\arabic{page}}
\pagebreak

\section{Introduction and summary of results}

According to AdS/CFT dictionary, 4-dimensional triangle anomaly of a global symmetry
maps to a 5-dimensional Chern-Simons term of the corresponding gauge symmetry \cite{Witten:1998qj}.
As the Chern-Simons term enters the equation of motion for the gauge field, one
is allowed to study dynamical aspects of triangle anomaly at strong coupling via holography.
A particularly interesting situation is finite temperature plasma with non-zero chemical potential,
which might be relevant in RHIC experiments. There has been a number of important studies on this aspect,
such as chiral magnetic effects \cite{Fukushima:2008xe,Kharzeev:2009pj,Buividovich:2009wi,Lifschytz:2009si,Yee:2009vw,Rebhan:2009vc}, hydrodynamic constitutive relations with triangle anomaly \cite{Son:2009tf,Lublinsky:2009wr}, etc to name a few examples.

In this work, we intend to study dynamical effects of triangle anomaly to long wave-length hydrodynamic
waves, focusing on their dispersion relations, via holography. Our aim is to analyze relevant linearized hydrodynamic modes in the background of charged AdS black-brane in the presence of 5D Chern-Simons coupling,
which is holographic dual to 4D chiral anomaly. Not to be too arbitrary, we choose to study
U(1) R-charged plasma of N=4 super Yang-Mills in field theory side, and for that we have a consistent 5D gauged supergravity truncation to the Einstein gravity plus U(1) Maxwell theory with a Chern-Simons term.
The action reads as
\be
(16\pi G_5){\cal L}= R +12 -{1\over 4}F_{MN}F^{MN} -{\kappa\over 4\sqrt{-g_5}}\epsilon^{MNPQR}A_MF_{NP}F_{QR}\quad, \label{5daction}
\ee
where $\kappa={1\over 3\sqrt{3}}$ for N=4 SYM, and we leave it arbitrary in our analysis.
An exact charged black-brane solution of Reisner-Nordstrom type for this theory is known, and has been used in holographic contexts for other purposes previously \cite{Son:2006em,Sin:2007ze,Erdmenger:2008rm,Torabian:2009qk,Edalati:2009bi}.
Note that N=4 super Yang-Mills in the presence of finite U(1) R-charge density is a {\it chiral} theory,
due to its $U(1)^3$ anomaly.

We first present complete equations of motion for linearized fluctuations out of the charged black-hole background, and classify them according to their helicity with respect to the $SO(2)$ rotation symmetry
transverse to the propagating direction. We adopt the Eddington-Finkelstein coordinate where simple {\it regularity} at the horizon is enough to implement the physical incoming boundary condition \cite{Yee:2009vw,Bhattacharyya:2008jc}.
We observe that the Chern-Simons term affects only the transverse shear modes of helicity $\pm 1$, that we will mainly concern about subsequently.
We develop a useful technique to solve the wave equations, and provide a systematic solution of the wave equations
in long wave-length expansion, which in principle allows one to proceed up to arbitrary higher order.
A concise formula for dispersion relations is obtained along the way, and we compute a few relevant lowest order terms explicitly, including the first non-trivial effect from anomaly.
Our result is
\be
\omega\,\, \approx\,\, -i{r_H^3\over 4m} \,\,k^2\,\, \pm\,\, i {\kappa Q^3\over 8m^2 r_H^3} \,\,k^3\,\, +\,\,{\cal O}\left(k^4\right)\quad:\quad{\rm helicity} \pm 1\quad,
\ee
where ${\cal O}(k^3)$ term (in fact any term with odd powers in $k$ ) is the new effect from chiral anomaly.
The dispersion relations are {\it chiral}, that is, the results depend on the sign of the helicity,
which can easily be expected from the chiral nature of N=4 SYM at finite R-charge density.
It will be interesting to obtain similar chiral dispersion relations from perturbative field theory calculations at weak coupling.
We leave analysis of other helicity modes, especially helicity 0 sound channel, to the future.

In fact, dispersion relations are in close relation to the constitutive relations in ref.\cite{Son:2009tf}, and one
should in principle be able to relate them. In this respect, our computation may also be considered as
providing one consistency check for the framework, as the method in ref.\cite{Son:2009tf} and ours are different from each other; ref.\cite{Son:2009tf} adopted non-linear fluid/gravity technique \cite{Bhattacharyya:2008jc,Rangamani:2009xk}, while ours is based on linearized quasi-normal analysis \cite{Policastro:2002se,Kovtun:2005ev,Son:2007vk}. They are two complementary techniques, although one needs to go one step further from the constitutive relations of the former to get dispersion relations from the latter.

We should also mention that while we are wrapping up our paper, there appeared a paper ref.\cite{Matsuo:2009xn} that
studies essentially same subject to ours. However, we find that our dispersion relations differ from theirs, especially the term of ${\cal O}(k^3)$, which we think should be present due to parity-breaking effects of the Chern-Simons coupling, but seems missing in ref.\cite{Matsuo:2009xn}.

\section{Complete linearized fluctuations of the charged AdS black-brane}

The equations of motion of the 5-dimensional theory (\ref{5daction}) consist of Einstein equation
\bear
{\cal E}_{MN} &\equiv& R_{MN}+\left(4+{1\over 12} F^2\right) g_{MN} -{1\over 2}F_{PM}F^P_{\,\,\,\,N} = 0\quad,
\eear
and the Maxwell equation of the U(1) gauge field,
\bear
{\cal M}_M &\equiv& \nabla_N F^{MN} +{3\kappa\over 4\sqrt{-g_5}}\epsilon^{MNPQR}F_{NP} F_{QR} =0\quad.
\eear
In our convention of cosmological constant in which asymptotic AdS space has unit radius,
the 5-dimensional Newton constant is explicitly given by
\be
G_5={\pi\over 2 N_c^2}\quad,
\ee
in relating to the dual N=4 super Yang-Mills theory.
The equations of motion allow an exact charged black-brane solution, which should holographically describe
a finite temperature plasma of N=4 SYM with non-zero U(1) R-charge chemical potential,
\bear
ds^2&=& -r^2 V(r)dt^2 +2 dt dr + r^2 \sum_{i=1}^3 \left(dx^i\right)^2\quad,\quad V(r)=1-{m\over r^4}+{Q^2\over 3 r^6}\quad,\nonumber\\
A&=&\left(-{Q\over r^2} +{Q\over r_H^2}\right)dt\quad,
\eear
where the horizon $r=r_H$ is located at the largest root of $V(r)=0$\footnote{Note that our normalization convention of the gauge field and the definition of $Q$ differ from those in previous literatures such as refs.\cite{Erdmenger:2008rm,Torabian:2009qk}, but it is easy to translate between them. }.
By studying small fluctuations of long wave-length in the above background, one can study {\it linearized} hydrodynamic waves of N=4 SYM at strong coupling in the presence of finite U(1) R-charge density.
See refs.\cite{Erdmenger:2008rm,Torabian:2009qk} for non-linear approaches via the technique of fluid/gravity correspondence.

Note that the above black-brane solution is written in Eddington-Finkelstein coordinate. As explained in ref.\cite{Yee:2009vw},
physical incoming boundary condition at the horizon can easily be achieved in this coordinate by requiring only {\it regularity} at the horizon, which makes solving the wave equations a lot simpler than the usual Poincare-type coordinate. Simply put, this is because the {\it regular} wave functions in Eddington-Finkelstein coordinate contain the necessary incoming-wave phase factor {\it automatically}.
We refer to the section 3 in ref.\cite{Yee:2009vw} for details.

Writing linearized fluctuations from the above background $(g^{(0)},A^{(0)})$ as
\be
g_{MN}=g^{(0)}_{MN}+\delta g_{MN}\quad,\quad A_M=A^{(0)}_M+\delta A_M\quad,
\ee
the first order variations of the equations of motion ${\cal E}_{MN}=0$ and ${\cal M}_M=0$
provide complete equations of motion for the fluctuations $(\delta g,\delta A)$.
Using diffeomorphisms and gauge transformations, one is allowed to work in the gauge where
\be
\delta g_{rr}=\sum_{i=1}^3 \left(\delta g_{ii}\right) = \delta A_r =0\quad,\quad \delta g_{ri}=0\quad,i=1,2,3\quad.
\ee
Since we are working directly with equations of motion, there is no issue of further constraints
for the gauge choice.
Also, it is important to remember that while we keep only first order variations in {\it magnitude} of fluctuations, we will keep arbitrary number of {\it derivatives} and do not invoke any derivative expansion.
This is in contrast to the non-linear fluid/gravity approaches where one truncates higher order derivatives at each step, while being non-linear in magnitudes. Both approaches seem to have their own pros and cons depending on the situation one is interested in.

From the variations of the Einstein equation, we have
\bear
\delta{\cal E}_{tt}&=& -{1\over 2}r^2 V\left(\partial_r^2\delta g_{tt}\right)
-{3rV\over 2}\left(\partial_r \delta g_{tt}\right)-{1\over 2 r^2}\left(\partial_i^2\delta g_{tt}\right)
+{3\over 2r}\left(\partial_t\delta g_{tt}\right)\nonumber\\
&+&r^2 V\left(\partial_t\partial_r\delta g_{tr} -{1\over 2}\partial_r\left(r^2 V\right)\partial_r\delta g_{tr}\right)+3rV\left(\partial_t\delta g_{tr}\right)-8r^2 V \delta g_{tr}\nonumber\\
&+& {1\over 2r^2}\left(2\partial_t\partial_i \delta g_{ti}-\partial_r\left(r^2V\right)\partial_i\delta g_{ti}\right)-{4QV\over 3r}\left(\partial_r \delta A_t\right)=0\quad,\\
\delta{\cal E}_{tr}&=& {1\over 2}\left(\partial_r^2\delta g_{tt}\right)+{3\over 2r}\left(\partial_r\delta g_{tt}\right)-\left(\partial_t\partial_r\delta g_{tr}\right)+{1\over 2}\partial_r\left(r^2V\right)\left(\partial_r\delta g_{tr}\right)\nonumber\\
&-& {1\over 2r^2}\left(\partial_i^2\delta g_{tr}\right)+8\delta g_{tr}+{1\over 2r^2}\left(\partial_i\partial_r\delta g_{ti}\right)+{4Q\over 3r^3}\left(\partial_r\delta A_t\right)=0\quad,\\
\delta{\cal E}_{rr}&=&{3\over r}\left(\partial_r\delta g_{tr}\right)=0\quad,\\
\sum_{i=1}^3\delta{\cal E}_{ii} &=& 3r\left(\partial_r\delta g_{tt}\right)+6\delta g_{tt}+3r^3V\left(\partial_r\delta g_{tr}\right)+24 r^2\delta g_{tr}
- \left(\partial_i^2\delta g_{tr}\right)\nonumber\\&+&\left(\partial_r\partial_i\delta g_{ti}\right)
+{4\over r}\left(\partial_i\delta g_{ti}\right)+{1\over r^2}\left(\partial_k\partial_i\delta g_{ki}\right)
-{2Q\over r}\left(\partial_r\delta A_t\right)=0\quad,\\
\delta{\cal E}_{ti}&=&-{1\over 2}\left(\partial_t\partial_i\delta g_{tr}\right)+{1\over 2r}
\partial_r\left(r^3 V\right)\left(\partial_i\delta g_{tr}\right)+{r^2V\over 2}\left(\partial_i\partial_r\delta g_{tr}\right)\nonumber\\
&+&{1\over 2}\left(\partial_r\partial_i\delta g_{tt}\right)+{1\over 2r}\left(\partial_i\delta g_{tt}\right)
-{1\over 2}\left(\partial_t\partial_r\delta g_{ti}\right)-{r^2V\over 2}\left(\partial_r^2\delta g_{ti}\right)-{rV\over 2}\left(\partial_r\delta g_{ti}\right)\nonumber\\
&+& 2V\delta g_{ti}+{1\over r}\left(\partial_t\delta g_{ti}\right)+{1\over 2r^2}\left(\partial_i\partial_j\delta g_{tj} -\partial_j^2\delta g_{ti}+\partial_t\partial_j\delta g_{ij}\right)\nonumber\\
&-&{Q\over r^3}\left(\partial_t\delta A_i -\partial_i\delta A_t +r^2 V\left(\partial_r\delta A_i\right)\right)=0\quad,\\
\delta{\cal E}_{ri}&=&
-{1\over 2}\left(\partial_r\partial_i\delta g_{tr}\right)+{3\over 2r}\left(\partial_i\delta g_{tr}\right)
+{1\over 2}\left(\partial_r^2\delta g_{ti}\right)+{1\over 2r}\left(\partial_r\delta g_{ti}\right)
-{2\over r^2}\delta g_{ti}\nonumber\\
&+&{1\over 2r^2}\left(\partial_r\partial_j\delta g_{ij}-{2\over r}\partial_j\delta g_{ij}\right)
+{Q\over r^3}\left(\partial_r\delta A_i\right)=0\quad,
\eear
and
\bear
\delta{\cal E}_{ij}-{1\over 3}\delta_{ij}\left(\sum_{k=1}^3 \delta{\cal E}_{kk}\right) &=&
-\left(\partial_i\partial_j-{1\over 3}\delta_{ij}\partial_k^2\right)\delta g_{tr}
+{1\over 2}\partial_r\left(\partial_i\delta g_{tj}+\partial_j\delta g_{ti}-{2\over 3}\delta_{ij}\left(\partial_k\delta g_{tk}\right)\right)\nonumber\\
&+&{1\over 2r^2}\left(\partial_i\partial_k\delta g_{kj}+\partial_j\partial_k\delta g_{ki}
-{2\over 3}\delta_{ij}\left(\partial_l\partial_k\delta g_{lk}\right)-\partial_k^2\delta g_{ij}\right)\nonumber\\
&+& {1\over 2r}\left(\partial_i\delta g_{tj}+\partial_j\delta g_{ti}-{2\over 3}\delta_{ij}\left(\partial_k\delta g_{tk}\right)\right) -\left(\partial_r\partial_t\delta g_{ij}\right)\nonumber\\
&+& {1\over 2r}\left(\partial_t\delta g_{ij}\right) -{1\over 2r}\partial_r\left(r^5V\partial_r\left({1\over r^2}\delta g_{ij}\right)\right) =0\quad.
\eear

From the Maxwell equation, one derives
\bear
\delta{\cal M}_t&=& {2Q\over r^3}\left(\left(\partial_t\delta g_{tr}\right)+r^2 V\left(\partial_r\delta g_{tr}\right) +{1\over r^2}\left(\partial_i\delta g_{ti}\right)\right)
- r^2V\left(\partial_r^2\delta A_t\right)\nonumber\\ &-&\left(\partial_t\partial_r\delta A_t\right)
-3rV\left(\partial_r\delta A_t\right)-{1\over r^2}\left(\partial_i^2\delta A_t\right)+{1\over r^2}\left(\partial_t\partial_i\delta A_i\right)=0\quad,\\
\delta{\cal M}_r&=&\left(\partial_r^2\delta A_t\right)+{1\over r^2}\left(\partial_i\partial_r\delta A_i\right)+{3\over r}\left(\partial_r\delta A_t\right)-{2Q\over r^3}\left(\partial_r\delta g_{tr}\right)=0\quad,
\eear
and lastly,
\bear
\delta{\cal M}_i&=&-2\left(\partial_t\partial_r\delta A_i\right)-r^2V\left(\partial_r^2\delta A_i\right)
-{1\over r}\partial_r\left(r^3 V\right)\left(\partial_r\delta A_i\right)-{1\over r}\left(\partial_t\delta A_i\right)\nonumber\\
&+& {1\over r^2}\partial_j\left(\partial_i\delta A_j-\partial_j\delta A_i\right)+\left(\partial_r\partial_i\delta A_t\right)+{1\over r}\left(\partial_i\delta A_t\right)
+{2Q\over r^3}\left(-\partial_r\delta g_{ti}+{2\over r}\delta g_{ti}\right)\nonumber\\
&+&{6\kappa Q\over r^4}\epsilon^{ijk}\left(\partial_j\delta A_k-\partial_k\delta A_j\right)=0\quad,
\eear
where the Chern-Simons coupling enters through the last term of the equation.
These are the complete equations of motion for linearized fluctuations from the background, which should be
useful in further analyses.

In studying waves with definite frequency $\omega$ and wave vector $\vec{\bf k}$, it is useful to organize the above equations of motion according to their helicity under the transverse SO(2) rotation which is a left-over symmetry once a wave vector is chosen. Without loss of generality, one may choose $\vec{\bf k}=k \hat{\bf e}_3$ resulting in a common phase factor
\be
e^{-i\omega t +i k x^3}\quad.
\ee
Equivalently, one simply replaces $\partial_t$ with $-i\omega$ and $\partial_i$ with $+i k \delta_{i3}$.
The equations are then classified into three categories; helicity $0$, $\pm 1$ and $\pm 2$.

The helicity $\pm 2$ modes are the simplest ones.
They are
\be
G_{\pm 2}\equiv {1\over r^2}\left(\delta g_{11}-\delta g_{22} \pm 2i \delta g_{12}\right)\quad,
\ee
 and satisfy the equations from $\left(\delta {\cal E}_{11}-\delta {\cal E}_{22} \pm 2i \delta {\cal E}_{12}\right)=0$, which takes a form of
\be
-{1\over 2r}\left(r^5 V(r)G_{\pm 2}'\right)' +i\omega r^2 G_{\pm 2}' +\left({3i\omega r\over 2}+{k^2\over 2}\right) G_{\pm 2} =0\quad,\label{h2eom}
\ee
where prime denotes derivative with respect to $r$. It is easy to derive shear viscosity by analyzing this equation, but it is not of our purpose here. On the contrary,
the most complicated channels are the helicity 0 ones. The modes involved are
\be
\delta g_{tt}\,,\quad \delta g_{tr}\,,\quad \delta g_{t3}\,,\quad
\delta g_{33}=-\left(\delta g_{11}+\delta g_{22}\right)\,,\quad \delta A_t\,,\quad{\rm and}\quad \delta A_3\quad,
\ee
which should satisfy the following ten linearized equations
\be
\delta {\cal E}_{tt}=\delta {\cal E}_{tr}=\delta {\cal E}_{rr}=\delta {\cal E}_{t3}=\delta {\cal E}_{r3}=\delta{\cal E}_{33}=\sum_{i=1}^3\left(\delta {\cal E}_{ii}\right)=\delta {\cal M}_t=\delta {\cal M}_r=\delta {\cal M}_3=0\,.
\ee
Although it is straightforward to write these equations explicitly from our previous equations of motion,
let us omit showing them explicitly as it is not of our particular interest in this paper.
One easily finds that Chern-Simons term does not play any role in helicity 0 and $\pm 2$ modes.
Related to that fact, the helicity $\pm 2$ equation (\ref{h2eom}) is independent of the sign of helicity.

Our present interest lies on the helicity $\pm 1$ modes of
\be
G_{\pm 1}\equiv {1\over r^2}\left(\delta g_{t1}\pm i\delta g_{t2}\right)\quad,\quad
H_{\pm 1}\equiv {1\over r^2}\left(\delta g_{13}\pm i\delta g_{23}\right)\quad,\quad
A_{\pm 1}\equiv \left(\delta A_{1}\pm i\delta A_{2}\right)\quad,
\ee
whose equations of motion are provided by
\be
\left(\delta{\cal E}_{t1}\pm i\delta{\cal E}_{t2}\right)=
\left(\delta{\cal E}_{r1}\pm i\delta{\cal E}_{r2}\right)=
\left(\delta{\cal E}_{13}\pm i\delta{\cal E}_{23}\right)=
\left(\delta{\cal M}_{1}\pm i\delta{\cal M}_{2}\right)=0\quad.
\ee
For subsequent uses, we write them down explicitly,
\begin{itemize}

\item $\left(\delta{\cal E}_{t1}\pm i\delta{\cal E}_{t2}\right)$ :
\be
-{V(r)\over 2r}\left( r^5 G_{\pm 1}'\right)'+{i\omega r^2\over 2} G_{\pm 1}'+{k^2\over 2} G_{\pm 1}
+{k\omega\over 2} H_{\pm 1} -{Q\over r^3}\left(-i\omega A_{\pm 1} +r^2 V(r) A_{\pm 1}'\right) =0\,,\label{eq1}
\ee

\item $\left(\delta{\cal E}_{r1}\pm i\delta{\cal E}_{r2}\right)$ :
\be
{1\over 2r^3}\left(r^5 G_{\pm 1}'\right)'+{ik\over 2}H_{\pm 1}' +{Q\over r^3} A_{\pm 1}' =0\quad,\label{eq2}
\ee

\item $\left(\delta{\cal E}_{13}\pm i\delta{\cal E}_{23}\right)$ :
\be
-{1\over 2r}\left(r^5 V(r)H_{\pm 1}'\right)'+i\omega r^2 H_{\pm 1}'+{3i\omega r\over 2}H_{\pm 1}
+{ik r^2\over 2} G_{\pm 1}' +{3ikr\over 2} G_{\pm 1} =0\,,\label{eq3}
\ee

\item $\left(\delta{\cal M}_{1}\pm i\delta{\cal M}_{2}\right)$ :
\be
-{1\over r}\left(r^3 V(r) A_{\pm 1}'\right)'+2i\omega A_{\pm 1}'+\left({i\omega\over r}+{k^2\over r^2}\mp {12\kappa Q k \over r^4}\right)A_{\pm 1} -{2Q\over r}G_{\pm 1}' =0\,,\label{eq4}
\ee
\end{itemize}
where one notices that the Chern-Simons coupling enters in the last equation with {\it opposite signs} for opposite helicities. As the Chern-Simons term mixes $A_1$ and $A_2$ in the above way, one is in fact {\it forced} to work in the present helicity basis.

Our task is to solve the above wave equations imposing suitable boundary conditions both at the horizon and the UV boundary $r\to\infty$. The physically sensible boundary condition at the horizon is the in-coming one,
for which regularity at the horizon is sufficient in Eddington-Finkelstein coordinate as discussed before.  At the UV boundary, we impose {\it normalizability} on the solutions to identify dynamical quasi-normal modes of the plasma \cite{Kovtun:2005ev}. This means practically that the solutions should vanish sufficiently fast near $r\to \infty$,
whose precise meaning will be discussed along the way.
Requiring the above two boundary conditions is a restrictive one, and gives us an {\it algebraic relation} between $\omega$ and $k$, which is called dispersion relation.
In other words, unless $\omega$ and $k$ satisfy a suitable dispersion relation, the mode can not be
accepted as a physical dynamical mode present in the plasma.

\section{Systematic solution of helicity $\pm 1$ shear modes}

In the present section, we provide a systematic solution of (\ref{eq1})-(\ref{eq4}) in long wave-length
expansion, based on the expectation that there exist hydrodynamic modes in the spectrum with the property
$\omega\to 0$ as $k\to 0$.
The philosophy is essentially same to the previous literatures, and differs from them only in the complexity of the present equations and a few technical details. Our novel result will be the solution given solely in terms of elementary integrals, which can in principle allows one to proceed up to arbitrary higher order in $\omega$ and $k$ without much difficulty.
There are in general other {\it massive} modes in the present linearized quasi-normal analysis (meaning $\omega \ne 0$ as $k\to 0$), which we don't intend to study at the moment.

One first finds that (\ref{eq3}) is in fact redundant given the first two equations (\ref{eq1}) and (\ref{eq2}). This is expected as we have four equations for three unknowns.
It is not straightforward to check this claim, but the way we show this is the following.
For a while, let us consider only helicity $+1$ modes omitting the subscript $+1$ for simplicity,
as the results for helicity $-1$ can easily be recovered by simple flipping of $\kappa\to -\kappa$.
From (\ref{eq1}), one can solve for $H$ algebraically in terms of $G$ and $A$, which facilitates solving the system significantly,
\be
H={2\over k\omega}\left({V(r)\over 2r}\left(r^5 G'\right)' -{i\omega r^2\over 2}G' -{k^2\over 2}G
+{Q\over r^3}\left(-i\omega A+r^2 V(r)A'\right)\right)\quad.\label{Heq}
\ee
Inserting this into (\ref{eq2}), one gets certain third order differential equation for $G$ and $A$.
Let's call this equation Eq.A. On the other hand, (\ref{eq2}) itself can be used to express $H'$
in terms of $G$ and $A$. Then, consider (\ref{eq3}) replacing $H'$ in the first two terms with the relation
one gets from (\ref{eq2}), and $H$ in the third term with (\ref{Heq}), to get another third order equation
for $G$ and $A$, which we call Eq.B. One can check that
\be
(\omega r^2)\cdot ({\rm Eq.A}) + k\cdot({\rm Eq.B})\equiv 0\quad,
\ee
which implies that (\ref{eq3}) is automatic once (\ref{eq1}) and (\ref{eq2}) are satisfied.

Therefore, one is left with only two differential equations to solve for $(G,A)$ by using (\ref{Heq}) for replacing $H$ in (\ref{eq2}), and the last equation (\ref{eq4}). After simple manipulations, they read as
\bear
\left({V(r)\over r}\left(r^5 G'\right)'\right)'+2Q \left({V(r)\over r} A'\right)'&=&
2i\omega r^2 G''+\left(7i\omega r+k^2\right) G'-{6i\omega Q\over r^4} A+{4i\omega Q\over r^3} A'\,,\nonumber\\
\left(r^3 V(r) A'\right)'+2Q G' &=& 2i\omega r A' +\left(i\omega +{k^2\over r}-{12\kappa Q k\over r^3}\right) A\,,\label{mastereq}
\eear
where we write them in such a way that terms with $\omega$ and $k$ are collected to the right-hand side.
Because $G$ appears only as $G'$, the above system is in fact second order differential equations for $(G',A)$. Once $(G,A)$ is found from the above, $H$ is finally given by (\ref{Heq}).

As we are interested in hydrodynamic modes, one invokes a series expansion
in $\omega$ and $k$ to solve them order by order iteratively,
\be
G(r)=\sum_{(m,n)\ge (0,0)}\,G^{(m,n)}(r) k^m \omega^n\quad,\quad
A(r)=\sum_{(m,n)\ge (0,0)}\,A^{(m,n)}(r) k^m \omega^n\quad.
\ee
Inserting the expansion in (\ref{mastereq}), the $(m,n)$-th order equations read as
\bear
\left({V(r)\over r}\left(r^5 G^{(m,n)'}\right)'\right)'+2Q \left({V(r)\over r} A^{(m,n)'}\right)'&=&
S^{(m,n)}\quad,\nonumber\\
\left(r^3 V(r) A^{(m,n)'}\right)'+2Q G^{(m,n)'} &=& T^{(m,n)}\quad,\label{expeq}
\eear
where the "sources" on the right, $S^{(m,n)}$ and $T^{(m,n)}$, arise solely from the lower order solutions than $(m,n)$. Explicitly they are
\bear
S^{(m,n)}&=& 2i r^2 G^{(m,n-1)''} +7i r G^{(m,n-1)'}+G^{(m-2,n)'}-{6iQ\over r^4} A^{(m,n-1)}+{4iQ\over r^3} A^{(m,n-1)'}\quad,\nonumber\\
T^{(m,n)} &=& 2ir A^{(m,n-1)'}+i A^{(m,n-1)} +{1\over r}A^{(m-2,n)} -{12\kappa Q\over r^3}A^{(m-1,n)}\quad,
\eear
and by definition $G^{(m,n)}=A^{(m,n)}=0$ when $m<0$ or $n<0$.
We will be able to solve (\ref{expeq}) in an integral form, so that one can solve the system order by order
by performing only elementary integrations, up to arbitrary order one desires.

One first integrates the first equation in (\ref{expeq}) to get
\be
\left(r^5 G^{(m,n)'}\right)'+2Q A^{(m,n)'} = {r\over V(r)} \int_{r_H}^r dr'\, S^{(m,n)}(r')\quad,\label{int1}
\ee
where we have fixed an integration constant to have a regular solution at the horizon $r=r_H$ where $V(r)=0$. The result is then used to replace $A^{(m,n)'}$ in favor of $G^{(m,n)}$ by
\be
A^{(m,n)'}= -{1\over 2Q}\left(r^5 G^{(m,n)'}\right)'+{r\over 2Q V(r)}\int_{r_H}^r dr'\, S^{(m,n)}(r')\quad,\label{rel1}
\ee
and inserting this to the second equation of (\ref{expeq}), one arrives at a second order differential equation of $G^{(m,n)'}$,
\be
\left(r^3 V(r)\left( r^5 G^{(m,n)'}\right)'\right)' -4Q^2 G^{(m,n)'} =
\left(r^4 \int_{r_H}^r dr'\, S^{(m,n)}(r')\right)'-2Q T^{(m,n)}\quad.
\ee
Surprisingly, this equation is in fact {\it integrable}, that is, the left-hand side can be transformed to a factorized form $A\left(B\left(C\cdot G^{(m,n)'}\right)'\right)'$ with
\be
A^{-1}=r^5 V'(r)\quad,\quad B=r^{13}V(r)\left(V'(r)\right)^2\quad,\quad C^{-1}=V'(r)\quad,
\ee
that facilitates subsequent integrations to solve for $G^{(m,n)'}$.
The method of finding such factorization was previously discussed in ref.\cite{Torabian:2009qk}; one first assumes
factorization in terms of yet unknown functions $(A,B,C)$ and finds that $C^{-1}$ is one {\it homogeneous}
solution of the differential equation. One can in fact easily generate at least one homogeneous solution via a simple coordinate transformation of the background solution, which in this case is
\be
t\to t+\epsilon x^1\quad,\quad x^1\to x^1+{\epsilon\over r}\quad,
\ee
giving $G_{homo}'=V'(r)$ and $A_{homo}'=-{2Q\over r^3}$. Once $C^{-1}$ is obtained, there is no further difficulty in determining $(A,B)$ by simple comparisons and integrations.

Defining
\be
U^{(m,n)}\equiv r^5 V'(r)\left[\left(r^4 \int_{r_H}^r dr'\, S^{(m,n)}(r')\right)'-2Q T^{(m,n)}\right]
\ee
in terms of $S^{(m,n)}$ and $T^{(m,n)}$, and integrating the above factorized form twice, one finally arrives at the useful integral form of the solution for $G^{(m,n)'}$,
\be
G^{(m,n)'}=V'(r)\int_{r_H}^r dr'\, {1\over (r')^{13} V(r') \left(V'(r')\right)^2} \int_{r_H}^{r'}dr''\,
U^{(m,n)}(r'')+C_1 V'(r)\quad,\label{final1}
\ee
where one integration constant has already been fixed for regularity at the horizon, while there remains
another integration constant $C_1$ free corresponding precisely to our previous homogeneous solution.
$A^{(m,n)'}$ is then given by (\ref{rel1}).
We won't need to present $G^{(m,n)}$ and $A^{(m,n)}$ as they are obtained from $G^{(m,n)'}$ and $A^{(m,n)'}$ unambiguously with boundary conditions specified before.
Notice that since $S^{(0,0)}=T^{(0,0)}=0$, the zero-th order solution $G^{(0,0)'}$ and $A^{(0,0)'}$
is simply given by our homogeneous solution, which means that $C_1$ can in fact be absorbed into a redefinition of the zero-th order solution. This will be assumed for all higher order solutions we present.

Although this completes the full iterative procedure for solving $G^{(m,n)'}$ and $A^{(m,n)'}$,
it is also useful to have a similar kind of integral form of $A^{(m,n)'}$ as for $G^{(m,n)'}$.
Starting instead from the second equation of (\ref{expeq}), one can replace $G^{(m,n)'}$ in terms of $A^{(m,n)'}$ by
\be
G^{(m,n)'}=-{1\over 2Q}\left(r^3 V(r)A^{(m,n)'}\right)'+{1\over 2Q} T^{(m,n)}\quad,
\ee
which is inserted into (\ref{int1}) to get a second order differential equation for $A^{(m,n)'}$,
\be
\left(r^5\left(r^3 V(r)A^{(m,n)'}\right)'\right)'-4Q^2 A^{(m,n)'}=\left(r^5 T^{(m,n)}\right)'-{2Qr\over V(r)}\int_{r_H}^r dr'\,S^{(m,n)}(r')\quad.
\ee
As expected, this is again integrable and the left-hand side can be rewritten as
$\tilde A\left(\tilde B\left(\tilde C\cdot A^{(m,n)'}\right)'\right)'$ with
\be
\tilde A^{-1} = V(r)\quad,\quad \tilde B=r^5 \left(V(r)\right)^2\quad,\quad \tilde C=r^3\quad.
\ee
Integrating twice, one gets to another useful integral form for $A^{(m,n)'}$,
\be
A^{(m,n)'}={1\over r^3}\int_{r_H}^r dr'\,{1\over (r')^5 \left(V(r')\right)^2}\int_{r_H}^{r'}dr''\,
V^{(m,n)}(r'')+{C_2\over r^3}\quad,\label{final2}
\ee
with a definition of
\be
V^{(m,n)}\equiv V(r)\left[\left(r^5 T^{(m,n)}\right)'-{2Qr\over V(r)}\int_{r_H}^r dr'\,S^{(m,n)}(r')\right]\quad.
\ee
Note that once we choose to have $C_1=0$, one has in general $C_2\ne 0$ which may be computed by comparing a couple of terms in the actual solution, but the specific value of $C_2$
will not be of our interest in subsequent sections.

To illustrate usefulness of our technique, we present a few lowest order solutions obtained by elementary integrations,
\bear
G^{(0,0)'}=V'(r)={4m\over r^5}-{2Q^2\over r^7} &,& A^{(0,0)'}=-{2Q\over r^3}\,,\nonumber\\
G^{(1,0)'}={6\kappa Q^3\over(2m r_H^2-Q^2)}\left({1\over r^5}-{r_H^2\over r^7}\right) &,& A^{(1,0)'}=-{6\kappa Q^2 r_H^2\over (2m r_H^2-Q^2)} {1\over r^3}\,,\nonumber\\
G^{(0,1)'}={-i V'(r)\over r_H^3}\int^r_{r_H} dr'\,
{r'\over V(r')}\left(1-{r_H^3\over (r')^3}\right)&,&A^{(0,1)'}=
{-iQ\over r_H^3 r^3}\left[\int_{r_H}^r dr'\left({1-{r_H^3\over (r')^3}\over V(r')}\right)' +{3 r_H\over V'(r_H)}\right].\nonumber
\eear

\section{Chiral dispersion relations}

From the solution in the previous section, one should be able to extract the dispersion relation
that $\omega$ and $k$ have to satisfy to fulfill boundary conditions. In-coming boundary condition as a regularity at the horizon has already been implemented along the procedure in the previous section.
What remains is the UV normalizability near $r\to\infty$, so we have to inspect asymptotic behaviors
of $(G,H,A)$ as $r\to\infty$. We first prove the following observation,

{\bf Proposition} :   $G^{(m,n)'}$ and $A^{(m,n)'}$ are at most ${\cal O}\left(r^{-3}\right)$  as $r\to\infty$. Moreover,
\be
G^{(m,n)'}\to \left[{1\over 2}\int_{r_H}^\infty dr' \,S^{(m,n)}(r')\right] {1\over r^3} + {\cal O}\left(r^{-5}\right)\quad.
\ee

{\it Proof} :
We prove it by induction. It is true for the zero-th order solution $G^{(0,0)'}=V'(r)$ and $A^{(0,0)'}={-2Q\over r^3}$ with $S^{(0,0)}=T^{(0,0)}=0$. Assuming it for $(m',n')<(m,n)$,
it is easy to see from the definitions that $S^{(m,n)}$ and $T^{(m,n)}$ are at most ${\cal O}\left(r^{-2}\right)$, so that the integral
\be
\int_{r_H}^\infty dr' \,S^{(m,n)}(r')\quad,
\ee
is finite. Then from the definition of $U^{(m,n)}$ and $V'(r)\to {4m\over r^5}$, one gets
\be
U^{(m,n)}\to 16m \left[\int_{r_H}^\infty dr' \,S^{(m,n)}(r')\right] r^3 +{\cal O}\left(r^2\right)\quad,
\ee
and from the integral form of $G^{(m,n)'}$, (\ref{final1}), it is straightforward to derive
\be
G^{(m,n)'}\to \left[{1\over 2}\int_{r_H}^\infty dr' \,S^{(m,n)}(r')\right] {1\over r^3} + {\cal O}\left(r^{-5}\right)\quad. \label{gasym}
\ee
Next, from the definition one easily finds $V^{(m,n)}\to {\cal O}\left(r^2\right)$, and from this one
derives that the integral
\be
\int_{r_H}^\infty dr'\,{1\over (r')^5 \left(V(r')\right)^2}\int_{r_H}^{r'}dr''\,
V^{(m,n)}(r'')
\ee
is finite. This, in conjunction with the integral form of $A^{(m,n)'}$ (\ref{final2}), gives one
\be
A^{(m,n)'}\to \left[\int_{r_H}^\infty dr'\,{1\over (r')^5 \left(V(r')\right)^2}\int_{r_H}^{r'}dr''\,
V^{(m,n)}(r'')+C_2\right]{1\over r^3} +{\cal O}\left(r^{-5}\right)\quad,
\ee
which completes the proof.

Remembering that $G\sim {\delta g\over r^2}$, the above means that $\delta g$ from $G$ is ${\cal O}(1)$,
which superficially looks ok with UV normalizability because the CFT metric perturbation is given by $\delta g^{CFT}\sim {\delta g\over r^2}$ which vanishes near $r\to\infty$ as ${\cal O}\left(r^{-2}\right)$.
As we will see shortly this is in fact a disguise, but the easiest way to find a problem with the UV normalizability is to look at $H$ simply given by (\ref{Heq}) once $(G,A)$ is found.
From our proposition above, one easily observes that
\be
H^{(m,n)}\to {1\over k\omega}\left[\int_{r_H}^\infty dr' \,S^{(m,n)}(r')\right]+{\cal O}\left(r^{-2}\right)\quad,
\ee
and because $H$ also corresponds to the CFT metric perturbation of $\left(\delta g^{CFT}_{13}+i\delta g^{CFT}_{23}\right)$, the above ${\cal O}(1)$ leading term is indeed non-normalizable.
The necessary and sufficient condition one has to impose for UV normalizability is then the algebraic relation
\be
\sum_{(m,n)\ge (0,0)} \left[\int_{r_H}^\infty dr' \,S^{(m,n)}(r')\right] k^m \omega^n =0\quad,
\ee
between $\omega$ and $k$, and this is our concise result for the dispersion relation.
We stress again that the procedure of computing the iterative solutions and the dispersion relation is completely in terms of integrations, which seems efficient to us.

Considering carefully the asymptotic behaviors of $(G,H)$, one can actually see a signal of non-normalizability already in the ${\cal O}\left(r^{-3}\right)$
term in $G^{(m,n)'}$. This is based on the general structure of near boundary solutions of Einstein equation with a negative cosmological constant, which has been studied before in the context of holographic renormalization \cite{de Haro:2000xn}.
According to it, a near boundary solution of ${\delta g\over r^2}$ has the structure
\be
{\delta g\over r^2}\sim \delta g^{CFT} +{\delta g^{(2)}\over r^2}+{\delta g^{(4)}\over r^4}+{\delta h^{(4)}{\rm log}r\over r^4}+\cdots\quad,
\ee
where $\delta g^{CFT}$ is a non-normalizable CFT metric perturbation, and $\delta g^{(2)}$
and $\delta h^{(4)}$ are completely determined by $\delta g^{CFT}$ through the Einstein equation of motion. The sub-leading part $\delta g^{(4)}$ contains CFT energy-momentum tensor which is not fixed by equations of motion.
Details are not essential here, but the important fact for us is that $\delta g^{(2)}$
and $\delta h^{(4)}$ {\it vanish} when the CFT metric is flat, which should be the case in the absence of
non-normalizable CFT metric perturbation. Therefore, ${\cal O}\left(r^{-3}\right)$
term in $G^{(m,n)'}$, which corresponds to $\delta g^{(2)}$, signals that there {\it must} be some non-normalizable perturbation to the CFT metric. In retrospect, this happens in $H^{(m,n)}$.
Because the problematic terms are proportional to the same factor in common,
\be
\left[\int_{r_H}^\infty dr' \,S^{(m,n)}(r')\right]\quad,
\ee
things seem to fit with each other.

From a few explicit lowest order solutions in the previous section, it is straightforward to
compute the relevant $S^{(m,n)}$'s as
\bear
&&S^{(0,0)}=S^{(1,0)}=0\quad,\quad S^{(0,1)}=-{12 i m\over r^4}\quad,\quad S^{(2,0)}=V'(r)\nonumber\\
&&S^{(1,1)}=-{18i\kappa Q^3\over (2mr_H^2-Q^2)}{1\over r^4}\quad,\quad S^{(3,0)}={6\kappa Q^3\over  (2mr_H^2-Q^2)}\left({1\over r^5}-{r_H^2\over r^7}\right)\quad,
\eear
from which one
finally arrives at
\bear
0&=& \sum_{(m,n)\ge (0,0)} \left[\int_{r_H}^\infty dr' \,S^{(m,n)}(r')\right] k^m \omega^n\\
&=&-{4im\over r_H^3} \omega + k^2 \mp {6i\kappa Q^3 \over r_H^3 (2m r_H^2-Q^2)}\omega k \pm {\kappa Q^3 \over 2 r_H^4 (2mr_H^2-Q^2)} k^3 +{\cal O}\left(\omega^2,\omega k^2,k^4\right)\,,\nonumber
\eear
where we recovered the sign dependence of helicity $\pm 1$.
This
is sufficient to solve for $\omega$ up to ${\cal O}\left(k^3\right)$ as
\be
\omega\,\, \approx\,\, -i{r_H^3\over 4m} \,\,k^2\,\, \pm\,\, i {\kappa Q^3\over 8m^2 r_H^3} \,\,k^3\,\, +\,\,{\cal O}\left(k^4\right)\quad,
\ee
where we have used $V(r_H)=0$ in the middle of calculations.
As the result depends on the sign of helicity, it features a chiral dispersion relation.
From the known thermodynamic quantities for this background
\be
\epsilon={3m\over 16\pi G_5}\quad,\quad p={m\over 16\pi G_5}\quad,\quad \eta={r_H^3\over 16\pi G_5}={s\over 4\pi}\quad,
\ee
the first term confirms the usual leading order piece from the shear viscosity \cite{Son:2007vk},
\be
\omega \sim -i{\eta\over (\epsilon+p)} k^2 + \cdots\quad,
\ee
while the next order term is the first new effect from anomaly.

\vskip 1cm \centerline{\large \bf Acknowledgement} \vskip 0.5cm

We thank Sang-Jin Sin for helpful correspondence.

 \vfil

\end{document}